\documentstyle[aps]{revtex}
\input{epsf}
\begin{document}
\draft
\twocolumn[\hsize\textwidth\columnwidth\hsize
           \csname @twocolumnfalse\endcsname

\title{Time dependence of current-voltage measurements of {\it c}-axis quasiparticle conductivity in 2212--BSCCO mesa structures}
\author{J.C. Fenton, G. Yang, C.E. Gough}
\address{Superconductivity Research Group, University of Birmingham, Edgbaston, Birmingham B15 2TT, United Kingdom.}
\date{\today}
\maketitle

\begin{abstract}
We report four-point IV measurements of the {\it c}-axis conductivity of
mesa structures of 2212--BSCCO, using a system with sub-$\mu$s resolution
along with multi-level pulses.
These allow a test to be made for the presence of nonequilibrium effects. Our results suggest simple heating alone is important in measurements of this kind.
\end{abstract}
]

\narrowtext

In the highly anisotropic high-temperature superconductor
Bi$_2$Sr$_2$CaCu$_2$O$_{8+\delta}$ (2212--BSCCO), a linear
array of intrinsic Josephson junctions is formed in the
out-of-plane direction due to the almost-insulating layers
separating the superconducting copper-oxide bilayers. To
investigate conduction across such junctions, mesa structures
can be fabricated by lithographic patterning on the
surface of almost atomically flat cleaved surfaces of
single crystals. Hysteretic multi-branched IV
characteristics are observed; each branch corresponds to
the expected irreversible characteristics of a junction in
the measured array.\cite{tink}

In principle, the shape of the characteristics can be used
to determine the {\it c}-axis quasiparticle conductivity and the
superconducting energy gap. However, in many circumstances
one observes regions in the IV characteristics with negative slope---``backbending''---and even an {\it S}-shaped
feature.  These cannot be described by simple
theory, but have been attributed by some authors to
nonequilibrium effects and by others to simple heating. In
the past, pulsed measurements on 200--500 ns time scales
have been used with this geometry in an attempt to
circumvent possible problems from sample heating. (See Ref.~\cite{jcf}
 for corresponding references.)

We have made pulsed four-point IV measurements on a stack of
(30 $\mu$m)$^2$ junctions with a system of cryogenically
cooled buffer amplifiers with 50 ns time resolution. Square voltage pulses are applied to the sample
via a series resistor. For more detail see Ref.~\cite{jcf}. Here we present data for an oxygen-annealed sample (Figs.~1 and 2) and for an as-grown sample from a different batch (Figs.~3 and 4).  Both samples have around 30 junctions in the main stack.

Typical IV measurements are
shown in Fig.~1. \cite{os} We identify the drop in
voltage with heating---the sample conductivity increases with temperature. Nonequilibrium effects are expected to vary on much shorter timescales.
Figure 2 shows a series of IV measurements at various times
after the switching-on of different-sized square current pulses. We believe the 50 ns
characteristics approximate the
intrinsic sample properties up to currents in excess of 20 mA. The 0.3 s characteristics are
indistinguishable from dc measurements.
The strong
time-dependence of the shape of the characteristics shows that the onset of the
backbending feature is not associated
with the gap voltage.
We deduce the temperature on the heated curves by
comparison with the 50 ns quasi-intrinsic characteristics
at higher base temperatures, e.g., for a current of
15 mA, the sample temperature rises from 33 K to above
T$_c$ in $\sim$ 10 $\mu$s.

\begin{figure}[!b]
\epsfxsize=8.5cm
  \epsfbox{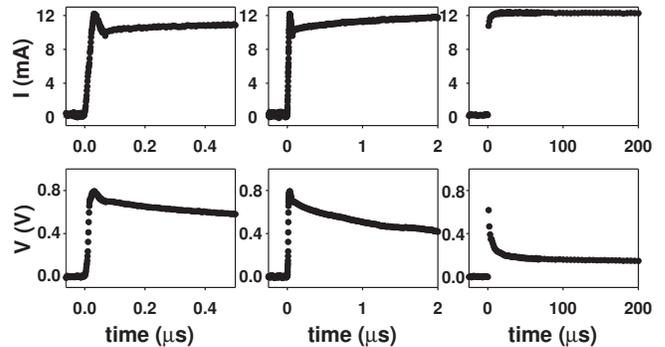}
\caption{Variation of sample current and voltage at 9 K for
an applied simple square pulse of 4 V.} \label{fig1}
\end{figure}
\begin{figure}[!t]
\epsfxsize=8.5cm
  \epsfbox{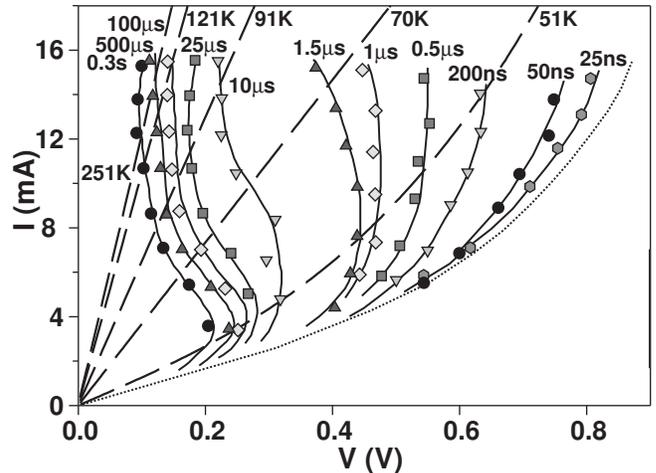}
\caption{Evolution of IV characteristics with time at a
base temperature of 33 K. Dashed lines show fits to 50 ns
IV characteristics at a number of base temperatures. Solid
lines are guides for the eye. The form of the intrinsic
characteristics at 33 K is suggested by the dotted line.}
\label{fig2}
\end{figure}
\begin{figure}[!b]
\epsfxsize=8.5cm
  \epsfbox{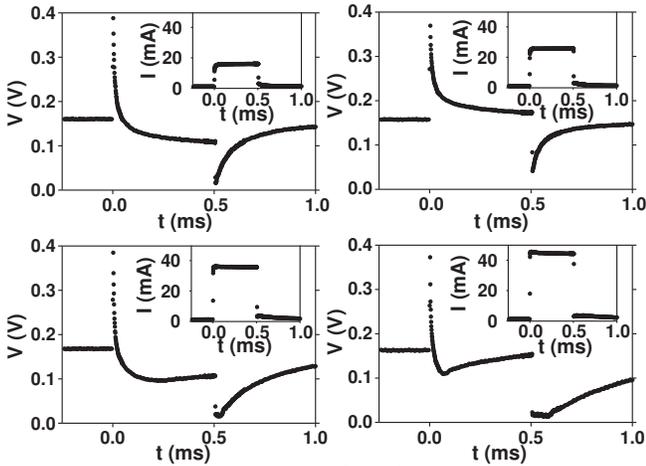}
 \caption{Sample voltage and (inset) current for pulses with a low applied level of 0.5V and 500 $\mu$s high applied levels of (a) 2V,(b) 3V,(c) 4V and (d) 5V, at a base temperature of 23 K. Note the small size of the current at low bias compared to its high-bias values.}
\label{fig3}
\end{figure}
Figure 3 shows measurements using high-bias pulse sequences superimposed on a non-zero low-bias level. The increase in low-bias sample voltage following the high-bias pulse is consistent with the mesa cooling back down---the low-bias conductivity decreases steeply from low temperatures towards T$_c$ and varies only slowly around and above T$_c$.
Examining the IV data in Figs.~3a--d, respectively 50 $\mu$s and 0.4 ms after the start of the high-bias pulse, we find backbending and an {\it S}-shaped feature.
In Figs.~3c and 3d, we note the transition from negative to positive slope, most likely associated with sample heating to well above T$_c$; this conclusion is supported by the initial regime of approximately constant voltage after the high-bias pulse, which corresponds to the mesa being heated above T$_c$.

The mesa temperature just after the end of the high-bias
pulse can be deduced from the low-bias sample voltage: this
is calibrated using measurements made at long times after
the high-bias pulse over a range of base temperatures.
In Fig.~4, the circle symbols show IV measurements at the end of a 5 $\mu$s high-bias pulse. We label these with the temperature inferred from the low-bias voltage shortly after the end of the high-bias pulse. The values are in good agreement with the temperature inferred from the position of the 5 $\mu$s IV characteristics with respect to the 50 ns characteristics.

These results underline the consistency of invoking a thermal origin for observed backbending and {\it S}-shaped features, without positing additional nonequilibrium effects. These measurements and their analysis in terms of thermal models give us confidence in interpreting the short-time IV characteristics, and will enable us to obtain more reliable information on the intrinsic interlayer tunnelling, free from the effects of heating.

This work was supported by the UK EPSRC.

\begin{figure}[!b]
\epsfxsize=8.5cm
  \epsfbox{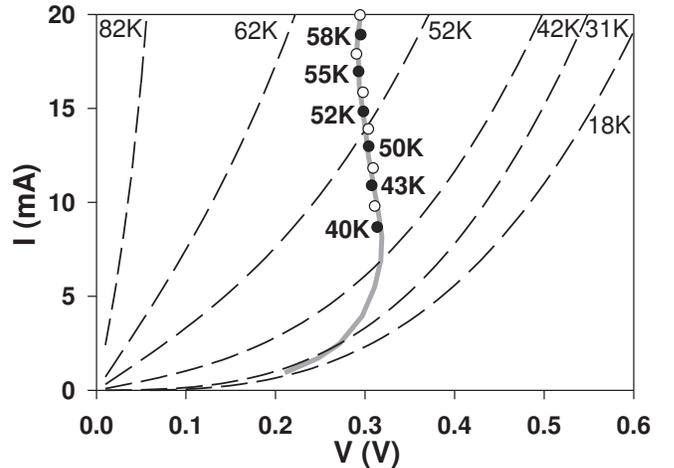}
\caption{IV characteristics just before the end of the 5
$\mu$s high-bias pulse (circle symbols), at a base
temperature of 18 K. The solid line is a guide for the eye.
The labels beside alternate points (filled circles)
indicate the temperature just after the end of the
high-bias pulse, inferred from the resistance of the stack
during the low level. Dashed lines show fits to 50 ns IV
characteristics at a number of base temperatures.}
\label{fig4}
\end{figure}

\end{document}